# Hunting micrometer-sized graphene flakes on gold substrate

Joel M. Katzen[†], Matěj Velický[†], Yuefeng Huang[†], Stacey Drakeley[†], William Hendren[†], Robert M. Bowman[†], Qiran Cai[‡], Ying Chen[‡], Lu Hua Li[‡], and Fumin Huang[†]*

[†] School of Mathematics and Physics, Queen's University Belfast, BT7 1NN, United Kingdom

[‡] Institute for Frontier Materials, Deakin University, Waurn Ponds, Victoria 3216, Australia

Gold is widely used as the substrate material in many graphene devices, due to its superior optoelectronic properties and chemical stability. However, there has been little experimental investigation on the optical contrast of graphene films on Au substrates. Here we report accurate measurement of the optical contrast spectra of few-layer graphene flakes on bulk Au. We used a high-resolution optical microscopy with a 100x magnification objective, accurately determining the thickness of flakes as small as one micrometer in lateral size, which are highly desired in many applications. The results are in excellent agreement with theoretical calculations and confirmed by independent Raman and AFM measurements. Furthermore, we demonstrate that the optical contrast spectroscopy is sensitive enough to detect the adsorption of a sub-monolayer airborne hydrocarbon molecules, which can reveal whether graphene is contaminated and opens the opportunity to develop miniaturized and ultrasensitive molecular sensors.

Graphene has attracted a lot of interest since its discovery in 2004[1], owing to a plethora of extraordinary performance in electronic, mechanical, thermal and optical properties. It is a zero-gap semiconductor due to the fact that the conduction and valence bands meet at the Dirac point[2], and has electron mobilities of more than 15000 $cm^2.V^{-1}.s^{-1}$ at room temperature[3]. As a result, it has the lowest resistivity (~$10^{-6}$ $\Omega.cm$) of any known conductive material at room temperature[4]. It also has exceptional mechanical strength and thermal conductivity[5]. These remarkable properties make graphene one of the most exciting materials widely investigated across multiple areas, including energy, electronics, photonics, and sensors[6,7]. For graphene devices, the choice of underlying substrate is critical and impacts the functionality of the device in many aspects, e.g., modulating the electrical and thermal conductivities, doping and affecting the Raman and fluorescence spectroscopy, and modifying the electrochemical properties[8]. Many applications require graphene to be deposited on metallic substrates, such as copper[9], nickel[10], silver[11] and gold[12-14], to achieve optimal electronic, plasmonic, catalytic and sensing functionalities[15-17]. Gold in particular is a popular choice of metallic material, due to its excellent optoelectronic properties and chemical stability. On the other hand, much of the research and applications require high quality pristine graphene films, which are often produced by mechanical exfoliation, resulting in small flakes of a few microns or even less. Small size films are also highly demanded in miniaturized nano-devices. Hence, accurate determination of the thickness of micrometer-sized graphene films on gold substrates is essential in graphene research and technology development.

Several methods are commonly used to identify the number of atomic layers in a graphene flake, including atomic force microscopy (AFM), Raman spectroscopy and optical contrast spectroscopy[18]. AFM is a low-throughput technique, taking up to several hours to produce a detailed enough image from which the important data can be extracted. It is also subject to the influence of a number of factors, such as humidity, surface roughness and contamination. Raman spectroscopy is much faster and can probe the flake's thickness non-destructively, based on the ratio of the G- and 2D-peaks, as well as the position and shape of the 2D-peak[19]. However, these spectral features can be influenced by many factors, including temperature[20], strain[21], and doping[22], therefore sometimes giving ambiguous results for the number of layers present. Compared to AFM and Raman spectroscopy, optical contrast spectroscopy has a number of distinct advantages: it is easy to implement, fast, noninvasive, and less affected by the aforementioned factors. As such, it has be-come an attractive technique for identifying the layer thickness of graphene and a wide variety of other 2D materials[23-25].

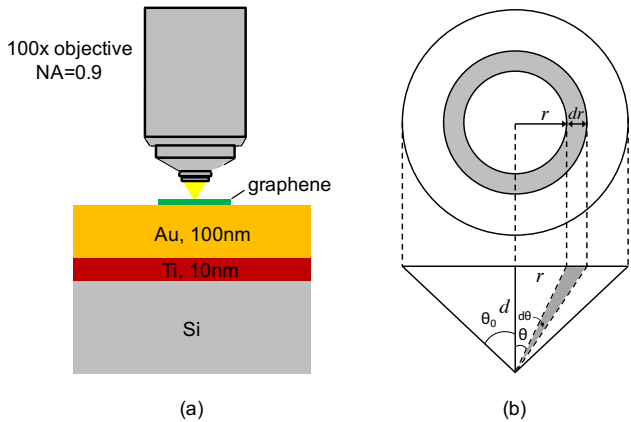

Figure 1 Schematic showing the experimental configuration (a) and the diagram of the incidence and focusing of light (b). The circle in the top panel of **b** refers to the incident aperture of the objective (top view). The triangle in the bottom panel indicates the focusing cone of light through the objective. The correspondence between the incident annular ring and the focusing cone is indicated by shaded areas.

Optical contrast of graphene films on oxidized silicon substrates has been studied extensively[26]. In contrast, very limited investigation has been carried out on metallic surfaces. Previously there was a theoretical investigation of the contrast of graphene on thin Au films, based on simple Fresnel reflectivity formulae at normal incidence[27]. However, there have been no experimental investigations of the optical contrast of graphene films on gold surfaces. Here we provide detailed experimental measurements, for the first time, of the optical contrast of few-layer (1-5 layer) graphene flakes on a bulk Au surface, using a high-resolution optical microscopy with a 100x magnification objective (Figure 1), which enables us to accurately identify the thickness of small flakes less than one micrometer lateral size.

The experimental results obtained are in excellent agreement with theoretical calculations from a modified model, taking into account the effects of different incident angles and polarizations, that become significant with the use of high numerical aperture objective. The accurate determination of the layer numbers was validated by independent AFM and Raman measurements. In addition, we found that the high accuracy and sensitivity of the optical contrast spectroscopy enabled the detection of the adsorption of a sub-monolayer contamination molecule. The optical contrast of freshly-prepared samples matched perfectly with that predicted by the theoretical model, whereas samples exposed to ambient air exhibited an increase of the contrast with time, as a result of the adsorption of airborne hydrocarbon molecules. This can reveal whether graphene is contaminated and could be exploited to develop miniaturized and highly sensitive optical sensors.

RESULTS AND DISCUSSION

Figure 1 shows the experimental configuration. Au film of 100 nm was deposited on Si/SiO$_X$ wafer through magnetron sputtering, with a 10 nm Ti adhesion layer (Figure 1). The Au film is thick enough to be equivalent to a bulk material. Less than 1.3% light can transmit through the Au film, so light reflected from the underlying substrate is negligible. To minimize the chance of contamination, graphene flakes were exfoliated onto Au films immediately after they were taken out of the sputtering chamber using the scotch tape method, and optical reflectance measurements were conducted soon afterward.

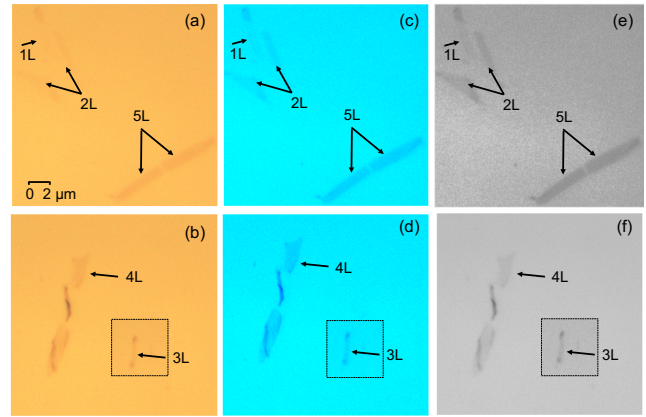

Figure 2 (a-b) Bright-field optical images of 1-5 layers graphene flakes without any optical filter, (c-d) with a 490 nm optical filter (10 nm bandwidth), and (e-f) greyscale images of (c-d). The contrast in (e) is enhanced by a factor of two for a better view of the monolayer film. The 3L film was at a different location on the sample and superimposed in the image.

Figure 2 shows the bright-field optical images of the exfoliated graphene films (1-5 layers, details about how the layer numbers are determined will be introduced later). Without using any optical filter, flakes of three layers and above are clearly distinguishable (Figure 2a-b), while the monolayer and bilayer flakes are only faintly visible. The contrast of the flakes is considerably enhanced when a 490 nm (10 nm bandpass width) band-pass filter is placed in front of the light source (Figure 2c-d). The bilayer flakes now are clearly visible. The visibility of the monolayer flake, though still faint, is also improved, which now can be visualized when the image is converted to greyscale and the contrast is enhanced by a factor of two (Figure 2e).

To accurately quantify the optical contrast of the graphene flakes, we measured the reflectance spectra on bare Au surfaces and on graphene flakes, respectively, and calculated the contrast spectra using the equation 3 (see MATERIALS AND METHOD). The results are shown in Figure 3a (red curves). It is evident that the contrast increases with the number of graphene layers. The maximum contrast occurs around 500 nm, which shifts very little between the 1 to 5 layer films. This explains why images filtered by a 490 nm optical filter show significantly enhanced optical contrast (Figure 2c-d).

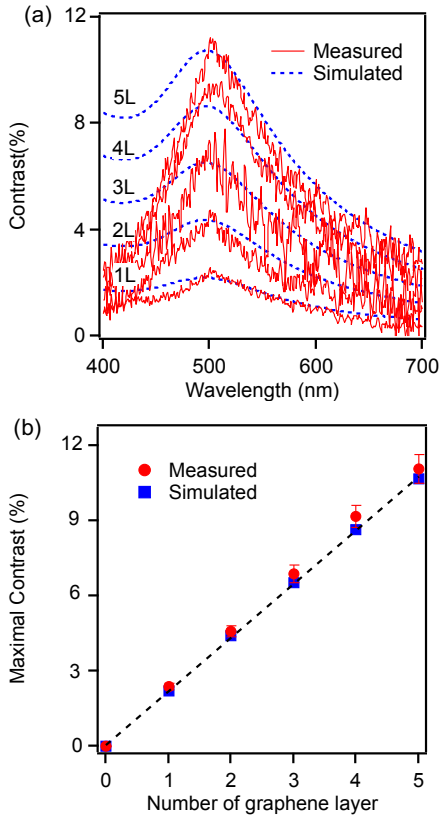

Figure 3 (a) Measured (red) and simulated (blue) optical contrast spectra of the 1-5 layers of graphene flakes shown in Figure 2. (b) The maximum optical contrasts of measured (red) and simulated (blue) results as a function of the number of graphene layers, fitted by a straight line passing through the origin point of zero contrast and zero thickness.

In previous research, simple Fresnel reflection formula at normal incidence is often used to calculate the reflection and the contrast spectra[26,27,29]. This produced reasonably good match with experimental results, through tailoring the refractive index (RI) of graphene, but resulting in various different RI values. Generally, reflectance is dependent both on the incident angle and the polarization of light. The difference may not be notable when the incident angle is small, such as in the case of a low magnification objective. However, the impact of the incident angle and polarization will become pronounced with increasing incident angle. The reflectance at large incidence angles could be significantly different from that of normal incidence. To illustrate this, we calculated the reflectance of the bare Au surface (no graphene) and the optical contrast of monolayer graphene on 100 nm Au (Figure 1a) at various incident angles and polarizations (Figure 4), based on the standard transfer matrix method[28]. The reflectance is almost constant up to 20°, but changes markedly with increasing incident angles (Figure 4a-b). As a result, the contrast is inhomogeneous across incident angles and polarizations (Figure 4c-d). In addition, the percentage of light incident at various angles is quite different. More light is incident at larger angles (Figure 1b). A high numerical aperture (NA) objective has a wide incident cone with a spanning angle of $\theta_0$ ($sin\theta_0$ =NA, e.g., NA=0.9, $\theta_0$ = 64°), as indicated in Figure 1b. It is thus necessary to appropriately average the contribution of various incident angles and polarizations to get the accurate reflectance and the contrast spectrum, which are given below (detail see Supporting Information):

$$R_{ave} = \frac{1}{2}(\bar{R}_{TE} + \bar{R}_{TM}) \quad (1)$$

$$\bar{R}_{TE,TM} = \frac{2\int_0^{\theta_0} R^{TE,TM}(\theta) \tan\theta (\sec\theta)^2 d\theta}{(\tan\theta_0)^2} \quad (2)$$

$\bar{R}_{TE,TM}$ is the averaged (over all incident angles) reflectance and $R^{TE,TM}(\theta)$ is the reflectance at the incident angle of θ, for TE and TM polarizations, respectively. The contrast is then calculated based on equation 3. The resulting calculated contrast spectra are shown in blue alongside the experimental data in Figure 3a. The thickness of graphene is given by d = 0.335N nm, where N is the number of layers. The refractive index of graphene was taken as 2.6-1.3$i$, the same as that of graphite, which has been shown to produce consistent simulation results with experiments[27,29]. The refractive index of Au is adopted from literature[30]. There is excellent agreement between the two sets of results in the long wavelength (λ> 500 nm) region, especially around the peaks. We plot the measured maximum optical contrasts of the flakes together with those simulated as a function of the layer number, shown in Figure 3b. As can be seen, the two sets of data match extremely well. The maximum contrast is linearly proportional to the number of layers, fitted very well by a straight line (dashed line). When the linear fitting curve is extrapolated, it passes through the origin point of zero contrast and zero thickness, as would be expected in an ideal situation. This confirms that the theoretical model is effective and the samples investigated were clean and of high quality. The discrepancy between the experimental data and the simulated results at short wavelength is notable. The calculated reflectivity spectrum perfectly matches the experimental one in the long wavelength range, but deviates considerably in the short wavelength range (Figure S2, Supporting Information). This is possibly due to a multitude of factors. Firstly, the assumption that graphene has a constant refractive index in the visible range is not true in reality. It varies with wavelength, and this dependence becomes more pronounced in the short wavelength region[31,32]. Secondly, the assumption there is equal amount of TE and TM polarization light may not be true either. Various optical elements (e.g., beamsplitter, fiber etc) may not function exactly with equal efficiency for TE and TM polarizations. The reflectivity of TE and TM polarizations deviates strongly in the short wavelength region. A slight imbalance between TE- and TM-polarized light will have a considerable impact on the overall reflectivity in the short wavelength range, but negligible in the long wavelength range (Figure S2). Some other factors may also contribute to the discrepancy (see more details in SI).

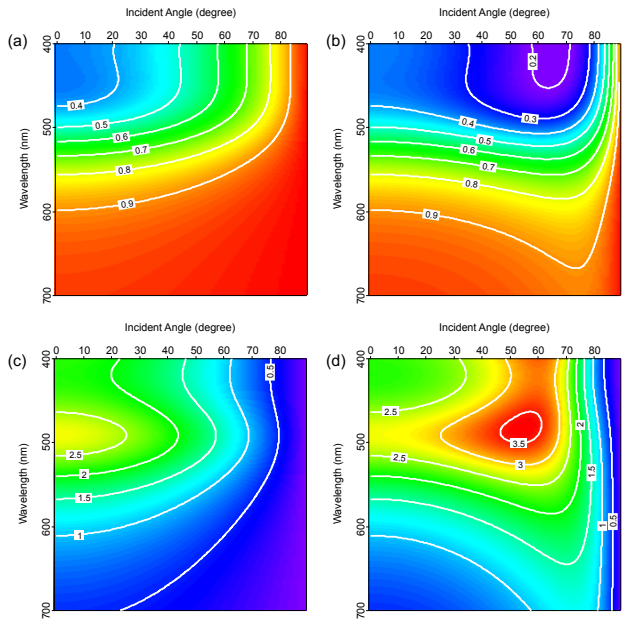

Figure 4 Calculated optical contrast and reflectivity of monolayer graphene on 100 nm Au, as functions of the incident angle and wavelength of light. (a-b) reflectance, (c-d) contrast. The left and right panels are for TE- and TM- polarizations, respectively, presented with same color scale. Refractive index of graphene $n_G = 2.6 - 1.3i$, thickness 0.335 nm.

To further confirm the accurate identification of the number of graphene layers, Raman and AFM measurements were carried out. The measured Raman spectra are shown in Figure 5a. As the thickness of graphene film increases, the G-peak becomes more intense in relation to the 2D-peak. Additionally, the 2D-peak gets broader, because on thicker flakes, the graphene film experiences a double resonance relating to intermediate phonon scattering electronic states[33]. The intensity of the G-peak scales linearly with the film thickness (Figure 5b). These properties are characteristic features of the Raman spectrum of few-layer graphene[19]. Here the G mode and the 2D mode of the monolayer graphene is at a similar intensity (Figure 5a). In normal circumstances, the 2D peak is stronger than that of the G peak[19]. However, it is known that the intensity of the 2D peak can be strongly affected by doping[34]. By contacting metallic substrates, the 2D peak can even be weaker than the G peak in monolayer graphene[22,35]. This shows the limitation of the Raman spectroscopy technique as a precise tool to accurately determine the graphene thickness.

Figure 6 shows the results from tapping-mode AFM measurements of the graphene flakes. Step profiles give heights of 0.56 ± 0.10, 0.75 ± 0.15, 1.1 ± 0.15, and 1.6 ± 0.2 nm for one, two, three, and five layers respectively. The measured thickness of two, three, and five-layer graphene flakes matched quite well with the expected values within experimental uncertainties. The height of the monolayer flake is slightly thicker than expected, which is not unusual in AFM measurements. AFM results are subject to the influence of a number of factors[36,37], such as the humidity of environment and the roughness of substrates. It has been observed experimentally that when graphene is placed on top of a rough surface, the flake acts like a tissue-sheet and follows the substrate's morphology[36,37]. When the substrate's roughness is higher than the flake's thickness, the measured height is deceptively higher than the flake's true height. It is thus not surprising that the measured height of the monolayer graphene film appears slightly thicker than expected. These results unambiguously confirmed that the number of the layers determined by the optical contrast method are accurate.

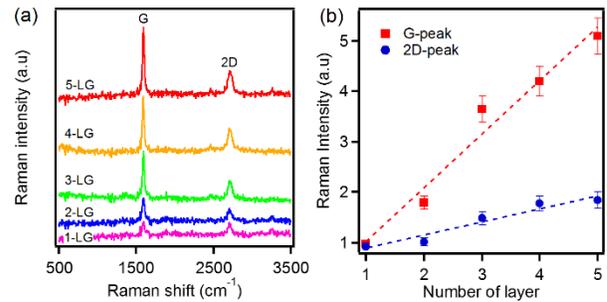

Figure 5 (a) Raman spectra measured for 1-5 layers of graphene, excited by a 532 nm laser. (b) Normalized Raman intensities of the G- (red) and 2D-peak as a function of the number of layers. The intensities of the G-peak and the 2D-peak of each flake are normalized to the intensities of the G- and 2D- peaks of the monolayer film.

Figure 3 reveals that graphene only has a modest contrast on bulk Au (~2.2% maximum contrast for monolayer), much lower than that on 90 nm SiO₂/Si substrates where the maximum contrast of monolayer graphene reaches up to 12%.[29] The low visibility of graphene on Au substrate is a result of the high reflectivity of Au surface (Figure S2), which is a downside shared by all reflective metals, such as Ag, Cu and Al.

Despite the modest optical contrast, the distinct contrast spectra of different layers and the linear dependency on the layer number make it a powerful tool to accurately identify the layer thickness of micrometer-sized graphene flakes, which are often generated by mechanical micro-cleavage method, to produce high quality pristine graphene. Compared to Raman and AFM measurements, the optical contrast method is fast, more definitive and less affected by factors such as strain, doping and temperature. In addition, the optical contrast of graphene is highly sensitive to the layer thickness, which changes drastically upon the addition of just one atomic layer of carbon atoms. Such a high sensitivity can have potential applications in many areas, for instance, to reveal whether graphene is contaminated, as when molecules are adsorbed onto graphene flakes, it will induce a notable change on the optical contrast spectrum and thus detectable. There is a wide variety of airborne molecules (e.g., alkanes, alkenes, aromatics, alcohol and water) in laboratory environment that could potentially adsorb onto graphene[38]. It is impractical to quantify every molecule species. To provide a guiding idea about how the adsorption of airborne molecules could impact the optical contrast of graphene, we simulated the optical contrast spectra of monolayer graphene on Au, with the adsorption of various thickness amorphous carbon film (Figure 7), which is closely relevant, as carbonaceous molecules are one of the major sources of airborne molecules.

avoided being significantly contaminated by dust. We repeatedly measured the optical contrast spectra after 2, 7, and 21 days, respectively. The results of the measured maximum positive contrast are shown in Figure 8a, which clearly demonstrate that the contrast of the exposed samples increased considerably in comparison to the original clean samples. This is more evident when we plot the contrast increments during different period of time intervals, i.e., 0-2 days, 2-7 days and 7-21 days, respectively, as shown in Figure 8b (the increment between 'x-y' days is the contrast difference between the 'y days' and the 'x days' in Figure 8a).

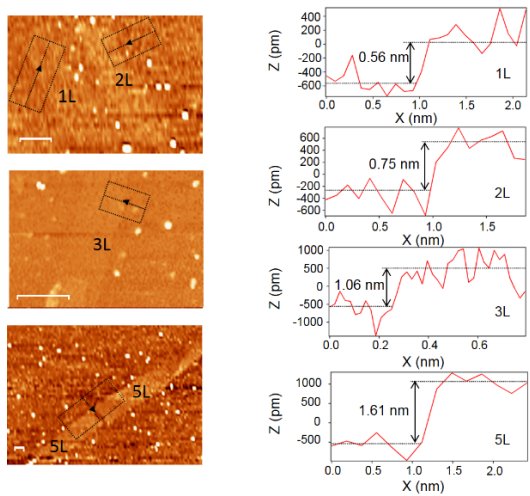

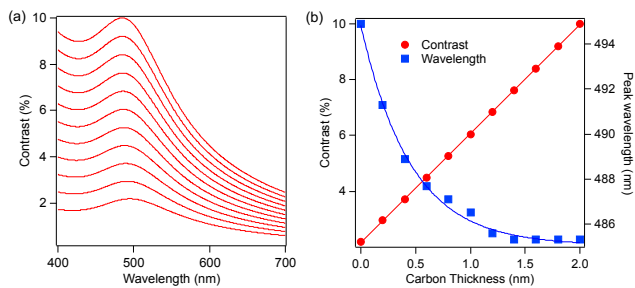

Figure 6 (Left) AFM images and (right) step profiles of (top-bottom) 1, 2, 3, and 5 layers of the graphene flakes shown in Figure 2. The height profiles are averaged within the boxes marked in the AFM images. Arrows indicate the directions of the height profiles. All scale bars correspond to 1 μm.

Figure 7 (a) Simulated contrast spectra of 0-2 nm (from bottom to top, 0.2 nm increment) amorphous carbon films adsorbed on monolayer graphene deposited on 100 nm Au. (b) The maximum contrast and the corresponding peak wavelength as a function of the thickness of amorphous carbon. The refractive index of amorphous carbon is adopted from ref.39.

Figure 7a demonstrates how the contrast spectrum evolves with the thickness of amorphous carbon (0 to 2 nm from bottom to top, at 0.2 nm incremental step). It is obvious the contrast increases with the thickness of amorphous carbon. The wavelength of the maximum contrast slightly blueshifts initially, but slows down with further increase of thickness. The trend can be visualized more clearly in Figure 7b, where the maximum contrast (circles) and the peak wavelength (squares) are plotted against the thickness of amorphous carbon. The maximum contrast increases linearly with thickness ($c = 3.9t + 2.17$, $t$ is the carbon thickness). The peak wavelength blueshifts following an exponential decay trend ($\lambda = 485 + 9.78e^{-2.2t}$). We can approximately estimate the sensitivity of the graphene system. The detection limit of the system is about 0.5%, estimated from the signal noise and experimental uncertainty level (Figure 3), which is equivalent to an average coverage of 0.13 nm amorphous carbon, about the thickness of one third monolayer graphene.

To test the idea of detecting airborne contamination molecules on graphene, we stored the samples in unsealed plastic boxes and kept them in ambient environment (initially the clean samples were kept in nitrogen-filled desiccators). In this way, the samples were allowed to contact air, but

The observed contrast increase is caused by the physisorption of airborne hydrocarbon molecules. Graphene and other 2D materials are known to adsorb airborne molecules in ambient environment, as demonstrated in wettability and electrochemical studies[38,40-45]. As shown in Figure 8b, initially the contrast increased notably after storage in ambient air for 2 days, after that the contrast increase slowed down considerably and almost plateaued after about one week, as indicated by the dashed guide line (blue). Such a time evolution trend is consistent with those reported in wettability studies[38,42,43], where the increase of the water contact angle exhibits a similar nonlinear behavior. The increments during the period of 7-21 days are small, below the detection limit (~0.5%), as the physisorption rate is drastically reduced after prolonged exposure in the air.[43]

A close examination of the contrast increment during the first 2 days reveals an interesting pattern: the contrast increment approximately scales with graphene thickness. Thicker films experienced larger increments, as indicated by the dashed red line. This is an intriguing observation, suggesting that the physisorption rate of airborne molecules on graphene at the early stage is not uniform: thicker films attracted more hydrocarbon molecules, therefore acquired a larger increment in contrast. If all graphene flakes had equal coverages of molecules, the contrast increment would be approximately the same for each film (for few layer graphene, the contrast increment is linear with increased layer thickness). The contrast of the '0~2 days' of the 3L film is unusually high.

This could be caused by sample drifting during the measurement, as the flake is quite small. If it were in the normal range (dashed box) guided by the red line, the increment of the 3L film during the period of '2~7' days would be in line with the trends of other flakes, as indicated by the dashed box.

Atomically-thin 2D materials are excellent absorbent for many molecules, which renders them promising platforms for sensors[44-45,48-49]. Molecules can attach to a solid surface by chemisorption or physisorption. Chemisorption is a strong interaction, usually involving specific chemical effects. In contrast, physisorption is a weak adhesion due to Van der Waals (vdW) interaction. Aromatic hydrocarbons can attach strongly to 2D materials due to strong π-π interaction, which was demonstrated to be stronger on monolayers than on multilayers[44]. Our results are consistent with recent investigations on the effect of airborne contamination molecules on the wettability property of graphene[38,43,46], in which the vdW interaction is shown to be the dominant mechanism, which increases with graphene thickness up to 4-6 layers[38,46].

Here we have demonstrated that the high sensitivity of the optical contrast spectroscopy allows to detect the adsorption of airborne molecules on graphene, hence can reveal whether graphene is contaminated, which is a piece of critical information central to a wide range of graphene applications, such as wettability, catalysis, adhesion, charge doping and carrier mobility. Graphene on Au substrate only has a modest contrast. The sensitivity can be significantly improved on high-contrast substrates, such as 90 nm $SiO_2$/Si substrates, which has a contrast gradient of 17.7%/nm for amorphous carbon on monolayer graphene (Figure S3). This will enable the detection of the adsorption of an average layer of 0.028 nm amorphous carbon (assuming the detection limit remains at 0.5%). Such a high sensitivity can be exploited to develop ultrasensitive molecular sensors. Graphene and a wide variety of 2D materials are newly-emerged platforms for sensors[45,48,49], with applications in broad areas, such as humidity, gas and protein sensing. Previously many graphene sensors are based on rather complicated protocols, relying on the measurements of electrical conductance or the Forster energy transfer in fluorescence[48,49]. The optical contrast method demonstrated here is a much simpler process, which is sensitive, fast, contactless, noninvasive, and can be ultimately miniaturized, as such, it provides an exciting new paradigm for sensors.

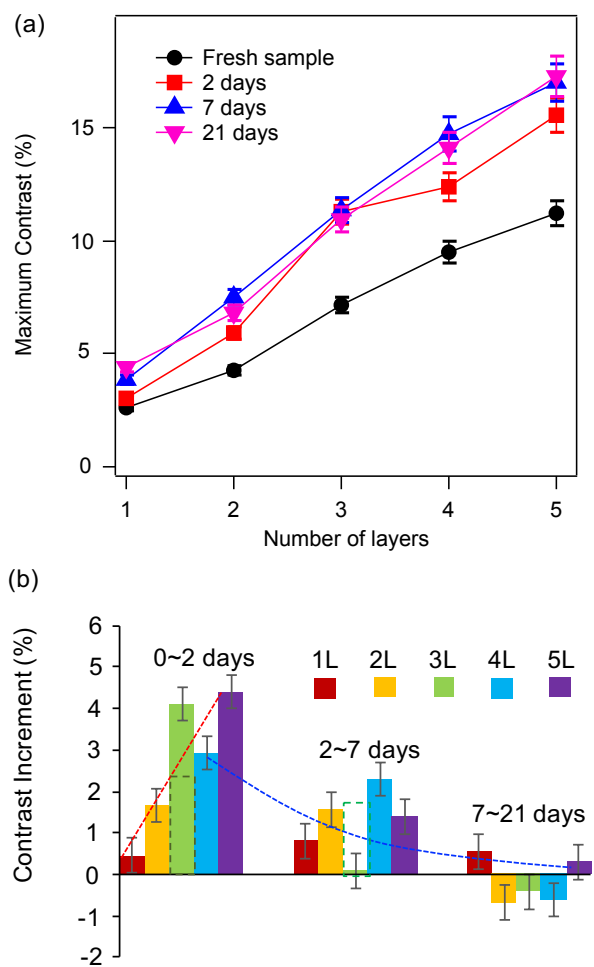

Figure 8 (a) Measured maximum optical contrast of freshly-prepared graphene flakes and after they were exposed in ambient air for 2, 7 and 21 days. (b) The increments of the optical contrast of the graphene flakes after exposure to air. Dashed lines are for guidance. Standard error bars are presented.

CONCLUSIONS

In summary, here we report a high-resolution optical contrast measurement of few-layer graphene films on bulk Au surface, using a 100x magnification objective. This provides a sub-micron spatial resolution, enabling the accurate identification of the thickness of graphene flakes less than one micrometer in lateral size, which is significantly relevant in research and applications requiring high-quality pristine graphene flakes and in the development of miniaturized graphene nanodevices. The optical contrast of graphene on bulk Au surface is much lower than that reported on oxidized silicon substrate, with a maximum optical contrast of about 2.2% for a monolayer graphene film (measured by an objective with NA=0.9). This is due to the high reflectivity of the gold surface, a feature shared by all reflective metals. The experimental results are found to match excellently with theoretically calculated results, when the effects of different incident angles and polarizations are appropriately taken into account. This is necessary when a high magnification objective is used. We further demonstrate that the high sensitivity of the optical contrast spectroscopy can detect the adsorption of a sub-monolayer airborne molecule and reveal whether graphene films are contaminated. This opens exciting opportunities for developing ultrasensitive molecular sensors.

MATERIALS AND METHODS

The samples were prepared by consecutively depositing 10 nm titanium and 100 nm gold onto Si/SiOx substrate by UHV magnetron sputtering. Graphene flakes were exfoliated onto the Au films immediately after they were taken out from the sputtering chamber and the optical contrast spectra were measured soon after the exfoliation. This minimizes the sample's exposure time to ambient air, mitigating the chance of contamination. The flakes were prepared using mechanical exfoliation from a natural graphite crystal (purchased from NGS Natur-graphit GmbH). It was cleaved with a high-tack, low-stain cello-tape to produce few-layer clean flakes, and pressed onto the freshly-sputtered gold surface before the tape was removed. Graphene flakes were identified with 100x Olympus objective lens (NA=0.9). Bright-field reflection spectra were measured using an Olympus microscope (model BX51) coupled to a QE65 Ocean Optics spectrometer, with the white light source of the microscope (12V 100W halogen lamp). An excitation laser of 532 nm with a power of 1 mW was used for Raman measurements. The signal was collected through a back-reflection configuration and coupled to a Jobin Yvon HR640 Raman spectrometer. AFM measurements were carried out in tapping mode on a Digital Instruments, Nanoscope IIIa. The data were processed with WSxM 4.0.[50] The step heights were analyzed with Nanoscope Analysis 1.5.

The optical contrast is calculated using the following equation:

$$C = 1 - \frac{R_{Gr}}{R_{Au}} \qquad (3)$$

where $C$ is the optical contrast, $R_{Au}$ is the reflectance of light on bare gold substrates, and $R_{Gr}$ is the reflectance of light on the Au substrate covered with a graphene flake. The contrast is positive (negative) when the addition of a graphene flake reduces (increases) the reflectance of light on Au substrate.

AUTHOR INFORMATION

**Corresponding Author**

* E-mail: f.huang@qub.ac.uk

**Author Contributions**

The manuscript was written through contributions of all authors. All authors have given approval to the final version of the manuscript.

ACKNOWLEDGMENT

This work was supported by the Department of Employment and Learning of Northern Ireland (DEL) and UK EPSRC (ref. EP/N025938/1)